\documentclass[12pt]{article}
\usepackage{authblk}
\usepackage{latexsym}
\usepackage{amsmath}
\begin{document}
\title{Black hole entropy as a consequence of excision}

\author{David Garfinkle\thanks{garfinkl@oakland.edu}}
\affil{Dept. of Physics, Oakland University,
Rochester, MI 48309, USA}
\affil{and Leinweber Center for Theoretical Physics, Randall Laboratory of Physics, University of Michigan, Ann Arbor, MI 48109-1120, USA}
\date{\today}
\maketitle

\begin{abstract}
Black hole entropy is shown to be a consequence of restricting our description of physics to the exterior of black holes.  This precludes the need for a statistical mechanical description of this entropy in terms of microstates.

\end{abstract}


A black hole event horizon represents a point of no return.  Nothing that happens inside the horizon can be viewed by observers outside the horizon, nor have any effect on anything outside the horizon.  
This fact has been used to great effect in numerical relativity in the technique of excision\cite{thornburgh,pretorius} where one simply removes the black hole interior from the computational grid.  However, excision can also be thought of as a conceptual device.  We can use our theories to describe only the physics of the exterior: the region outside of all black hole event horizons.  This has the advantage that we don't expend effort making predictions of things we can't observe.  However, it seems that the physics of the exterior lacks conservation laws since things fall into the black hole and are thus lost from the exterior.  But we can rescue conservation laws by using Gauss's law and its analogs.  Recall that Gauss's law says that the charge enclosed by a surface is equal to the integral of (the normal component of) the electric  field over that surface divided by $4\pi$.  Thus, we can {\it call} $1/(4\pi)$ times the integral of the electric field over the black hole event horizon {\it the charge} of the black hole.  Our physics of the exterior now has a law of conservation of charge: any charge lost by the exterior shows up as charge ``of the black hole'' as computed by this integral.   
In the language of differential forms: the electromagnetic field is described by a two-form $\bf F$, and the charge and current density is described by a one-form $\bf j$.  Then one of Maxwell's equations is $d *{\bf F}= 4 \pi *{\bf j}$ which leads to integral over a volume of $*{\bf j}$ is equal to integral over its boundary surface of $*{\bf F}/(4\pi)$.  So we call integral of $*{\bf F}/(4\pi)$ over the horizon the charge of the black hole. Similar considerations apply to conservation of energy: for stationary spacetimes there is a timelike Killing vector $\boldsymbol \xi$ and a relation between the integral of $*d{\boldsymbol \xi}/(8\pi)$ over a surface and the energy contained in the volume bounded by that surface.  So we can {\it call} integral of $*d{\boldsymbol \xi}/(8\pi)$ over the horizon the energy of the black hole.  

To put it another way, integral of $*{\bf F}/(4\pi)$ is a ``charge-like quantity'' that takes the place, in our physics of the exterior, of the charge inside the black hole. Similarly integral of $*d{\boldsymbol \xi}/(8\pi)$ is an ``energy-like quantity'' that takes the place of the energy inside the black hole.

In addition to charge and energy, black holes have a tempertature. As shown by Hawking\cite{hawking3} in his celebrated calculation, black holes radiate a thermal spectrum of particles at temperature $T=\kappa /(2\pi)$ (in units where $G, \, c, \, \hbar$ and Boltzmann's constant are equal to 1).  Here $\kappa$ is the surface gravity of the black hole given by the relation that for the Killing vector $\chi ^a$ normal to the black hole, we have ${\chi ^a}{\nabla _a}{\chi ^b} = - \kappa {\chi ^b}$ on the horizon.  

Using the first law of thermodynamics (a relation between temperature, energy, and entropy) the result $T=\kappa /(2\pi)$ leads to the relation $S=A/4$ where $S$ is the black hole entropy and $A$ is the area of the black hole event horizon.  This identification of black hole area with entropy is made even stronger by (the second law of thermodynamics and) the Hawking area 
theorem\cite{hawking2} which states that the area of black hole event horizons increases. Indeed, even before the result of\cite{hawking3}, Bardeen, Carter and 
Hawking\cite{hawking1} found a ``law of black hole mechanics'' which formally looks like the first law of thermodynamics, and Bekenstein\cite{bekenstein} had conjectured that black holes have an entropy proportional to event horizon area.  

But the laws of thermodynamics are themselves consequences of statistical mechanics, in which 
$S = \ln W$ (in units where Boltzmann's constant is 1) where $W$ is the number of microstates compatible with the observed macrostate of the system.  Thus the claim that $S=A/4$ for black holes seems to be the claim that a black hole of area $A$ has $e^{A/4}$ microstates.  But a precise definition of microstate number is essentially the dimension of the subspace of the quantum Hilbert space compatible with the macrostate.  Thus a correct calculation of black hole microstates should be done in a theory of quantum gravity.  Since we know that the answer is supposed to be $e^{A/4}$, one can view this calculation as a test that a correct quantum theory of gravity needs to pass.

And indeed there are calculations of the number of black hole microstates done in string theory\cite{andy} and in loop quantum gravity.\cite{abhay}  However, the fact that there are {\it two} calculations should itself give us pause: string theory and loop quantum gravity can't {\it both} be the correct theory of quantum gravity.  And why would an {\it incorrect} theory of quantum gravity {\it just happen} to give the correct answer for the number of black hole microstates?  It seems to me that these calculations tell us less about black hole microstates than they tell us about the current state of our theories of quantum gravity: namely that these theories are sufficiently vague about precisely {\it which} formal calculation in the mathematics of the theory corresponds to a given physical effect.  Thus, if there is a physical question for which the answer is already known and a formal calculation in a theory of quantum gravity which purports to obtain that answer, it is my opinion that the formal calculation should be taken with a (very large) grain of salt. 

There are also awkward questions about exactly {\it where} these microstates are located: inside the black hole or on the horizon? As shown in\cite{tedetal} either answer to this question leads to difficulty.  Since one could pose similar questions about where the black hole charge is located,  by analogy perhaps the question about black hole microstates could be resolved by thinking of black hole area as an ``entropy-like quantity'' rather than an entropy.

The rest of this Note will spell out this alternate point of view on black hole entropy.  It seems to me that black hole thermodynamics is very different from ordinary thermodynamics: In ordinary thermodynamics, the fundamental quantity is entropy, which is given in terms of the logarithm of the number of microstates.  Temperature is a derived concept having to do with how entropy depends on energy.  In black hole thermodynamics temperature is the fundamental quantity, and entropy is a derived quantity that has to do with how temperature depends on black hole energy.  

In ordinary thermodynamics thermal states arise because different parts of the system can exchange energy and do so in such a way as to maximize entropy.  But black holes are one-way systems: the interior cannot emit anything into the exterior.  Why, then do black holes radiate a thermal spectrum?  The answer has to do with particular properties of quantum field theory and of black holes.  It is to be expected that whatever its initial state, eventually the quantum state (of whatever matter fields there are) should settle down to a stationary state.  In a stationary state, one would expect steady accretion on to the black hole.  The surprising thing is that it is a steady accretion of {\it negative} energy (which tends to shrink the black hole) along with a compensating outward flow of positive energy.  For the case of static black holes one can 
see\cite{hawking4} that this outward flow is thermal as follows: the Euclidean version of a static black hole must be periodic in time in order to be smooth at the horizon.  But in Euclidean quantum field theory, quantum states that are periodic in time must be thermal states, and there is a relation between the temperature and the period in Euclidean time.  Things are a little more complicated for black holes that are stationary but not static.  Nonetheless, one can 
show\cite{hawking4,wald1} that in the presence of a stationary black hole, a stationary quantum field state must be a thermal state, and that Hawking's formula relating temperature to surface gravity holds. 

Given the temperature and energy of the black hole, it is a property not just of general relativity but of any diffeomorphism-invariant theory\cite{wald2} that there is an entropy-like quantity $S$ whose relation to temperature and energy are the same as the relation given by the usual first law of thermodynamics.  Furthermore, in general relativity this entropy-like quantity is given by $S=A/4$ where $A$ is the black hole horizon area.  One is free to call this entropy-like quantity ``the black hole entropy'' but note that there is {\it no implication} that this ``entropy'' comes from a logarithm of black hole microstates as an entropy in ordinary thermodynamics would.

Most importantly, we may have seriously overestimated the extent to which black hole thermodynamics can provide guidance in finding a quantum theory of gravity.  It seems that black hole thermodynamics is a self-contained body of knowledge dependent only on the properties of quantum fields in the presence of black holes.  It thus has no need to be derived from a more fundamental theory, and therefore provides very little in the way of clues as to what that more fundamental theory might be.  This is a sobering conclusion, but if it is true then at least we can better direct our efforts by ceasing to look in the wrong directions.    


\section*{acknowledgement}
I would like to thank Bob Wald for helpful discussions.  This work was supported by NSF grant numbers PHY-1505565 and PHY-1806219 to Oakland University.

\end{document}